\documentclass[a4paper,11pt]{article}
\usepackage{pos}

\newcommand{\be}{\begin{equation}}
\newcommand{\ee}{\end{equation}}
\newcommand{\bea}{\begin{eqnarray}}
\newcommand{\eea}{\end{eqnarray}}
\newcommand{\ds}{{\sf DarkSUSY}}
\newcommand{\code}[1]{{\tt #1}}

\title{DarkSUSY 6.3 -- Freeze-in, out-of-equilibrium freeze-out, cosmic-ray upscattering and further new features}
\ShortTitle{DarkSUSY 6.3 -- Freeze-in and more}

\author*[a,b]{Torsten Bringmann}
\author[c]{Joakim Edsjö}

\affiliation[a]{Department of Physics, University of Oslo, Box 1048, N-0316 Oslo, Norway}
\affiliation[b]{Theoretical Physics Department, CERN, 1211 Geneva 23, Switzerland}
\affiliation[c]{ The Oskar Klein Centre for Cosmoparticle Physics, Department of Physics, 
   Stockholm University, AlbaNova, SE-106 91 Stockholm, Sweden.}

\emailAdd{torsten.bringmann@fys.uio.no}
\emailAdd{edsjo@fysik.su.se}

\abstract{
DarkSUSY is a versatile tool for precision calculations of a large variety of dark matter-related 
signals, ranging from predictions  for the dark matter relic density to dark matter self-interactions 
and rates relevant for direct and indirect detection experiments. 
In all of these areas significant new code additions have been made in recent years, since the release of 
DarkSUSY 6 in 2018, which we summarize in this overview. In particular, DarkSUSY now allows users to 
compute the relic density for feebly interacting massive particles via the freeze-in mechanism, but 
also offers new routines for freeze-out calculations in the presence of secluded dark sectors 
as well as for models where kinetic equilibrium is not fully established during the freeze-out process.
On the direct detection side, the effect of cosmic-ray upscattering of dark matter has been fully
implemented, leading to a subdominant relativistic component in the expected dark matter flux at Earth.
Finally, updated yields relevant for indirect searches with gamma rays, neutrinos or charged
cosmic rays  have been added; the new default spectra are based on a large number of 
{\sf Pythia 8} runs, but users can also easily switch between various alternative spectra.
Further code details, including a manual and various concrete example applications, are provided at 
{\tt  www.darksusy.org}.
\vspace*{-17cm}\mbox{ }\\
\flushright{CERN-TH-2022-031}
\vspace*{16cm}
}

\FullConference{%
  Computational Tools for High Energy Physics and Cosmology (CompTools2021)\\
  22-26 November 2021\\
  Institut de Physique des 2 Infinis (IP2I), Lyon, France
}

\renewcommand{\logo}{\relax}

\tableofcontents

\begin{document}
\maketitle

\section{Introduction}

The identity of dark matter (DM) remains one of the central questions of fundamental physics,
even though its present abundance of $\Omega_\chi h^2=0.12$ has been 
precisely measured over an impressive range of cosmological distance scales~\cite{Planck:2018vyg}.
The minimal requirement on any theory beyond the standard model (BSM) that includes a potential DM 
candidate is therefore a DM production mechanism in the early universe explaining the observed value of $\Omega_\chi h^2$.
Further experimental observables, based on non-gravitational interactions introduced in such 
BSM theories, are generally needed in order to actually close in on the nature of DM. 
Consequently, there is a huge demand on precision calculations both for the relic density 
and rates associated to other potentially observable DM signals. \ds~\cite{ds4,Bringmann:2018lay} 
is one of the major public and generic codes to perform such calculations 
(complemented by {\sf MicrOMEGAs}~\cite{Belanger:2018ccd} and {\sf MadDM}~\cite{Ambrogi:2018jqj}, 
which each have a somewhat different focus~\cite{Arina:2020alg}).

The upgrade to \ds~6~\cite{Bringmann:2018lay} represented a major overhaul and restructuring
of the code. The main novelty introduced in that release, besides many new physics features, 
is a highly modular structure that allows users to numerically compute DM properties beyond supersymmetric 
models (and more generally beyond weakly interacting massive particles as DM candidates).
Since then, \ds~has been widely used in the community. Recent major applications include sensitivity 
studies for a DM signal from the Galactic center by the Cherenkov Telescope Array (CTA) collaboration~\cite{CTA:2020qlo},
searches for DM-related gamma-rays from the sun by the HAWC~\cite{HAWC:2018szf} and Fermi~\cite{Serini:2020yhb} 
collaborations, as well as searches for neutrino signals with Super-Kamionkande~\cite{Super-Kamiokande:2020sgt} 
and IceCube~\cite{Tonnis:2021krs}; furthermore, \ds\ is one of the main backends that 
{\sf DarkBit}~\cite{GAMBITDarkMatterWorkgroup:2017fax} relies on, for its rate and relic density calculations,
and as such decisive for the global fits performed by the {\sf GAMBIT} collaboration~\cite{GAMBIT:2017yxo,Bloor:2021gtp}.
At the same time, the code has seen further active development and new features added, warranting
an updated description beyond what is regularly reported on the homepage.\footnote{%
{\tt  http://www.darksusy.org}
}

In these proceedings we summarize the most  important updates since version 6.1.~\cite{Bringmann:2018lay} of the code,
both in terms of physics and actual implementation.\footnote{%
These proceedings do however {\it not} replace Ref.~\cite{Bringmann:2018lay} as the correct way of 
referring to the most recent version of \ds~6. If you use \ds, please also consider citing
Ref.~\cite{ds4} for code prior to version 4.2 that is still contained in the current release. 
Finally, most routines in \ds\  have been implemented 
in the context of original research work. Therefore, when using those routines, 
please also give proper credit to the relevant articles indicated in section 5 of the manual (as well
as in the respective sections in these proceedings).
}
The text is organized along the three main directions where significant code updates
have been implemented, namely relic density calculations 
(section \ref{sec:rd}),  direct detection (section \ref{sec:crdm}) and indirect detection
routines (section \ref{sec:id}). After a brief summary, in section \ref{sec:sum}, we 
also include a more technical Appendix \ref{app} where we describe recent updates to 
the installation and make system that address in particular commonly encountered 
problems when building contributed code like {\sf HEALPix}.

\section{Relic density: Boltzmann equations}
\label{sec:rd}
The evolution of the DM phase-space density $f_\chi(t,p)$ in the early universe
is governed by the Boltzmann equation
\be
  \label{diff_boltzmann}
  E\left(\partial_t-Hp\partial_p\right)f_\chi=C_{\rm ann}[f_\chi] + C_{\rm el}[f_\chi]\,,
\ee
where
\bea
  \label{Cand_def}
  C_\mathrm{ann}&=&\frac{1}{2g_\chi}\int\frac{d^3\tilde p}{(2\pi)^32\tilde E}\int\frac{d^3k}{(2\pi)^32\omega}\int\frac{d^3\tilde k}{(2\pi)^32\tilde \omega}(2\pi)^4\delta^{(4)}(\tilde p+p-\tilde k-k) \left|\mathcal{M}\right|^2_{\chi\chi\leftrightarrow \psi\psi}\nonumber\\
&&\times
\left[
f_\psi(\omega)f_\psi(\tilde \omega)\bar f_\chi(E) \bar f_\chi(\tilde E) 
-f_\chi(E)f_\chi(\tilde E) \bar f_\psi(\omega) \bar f_\psi(\tilde \omega)
\right]\,,\\
  \label{Cel_def}
  C_\mathrm{el}&=&\frac{1}{2g_\chi}\int\frac{d^3\tilde p}{(2\pi)^32\tilde E}\int\frac{d^3k}{(2\pi)^32\omega}\int\frac{d^3\tilde k}{(2\pi)^32\tilde \omega}(2\pi)^4\delta^{(4)}(\tilde p+p-\tilde k-k)  \left|\mathcal{M}\right|^2_{\chi\psi\leftrightarrow \chi\psi}\nonumber\\
&&\times
\left[
f_\chi(E) f_\psi(\omega) \bar f_\chi(\tilde E)  \bar f_\psi(\tilde \omega)
-f_\chi(\tilde E) f_\psi(\tilde \omega) \bar f_\chi(E) \bar f_\psi(\omega)
\right]
\eea
describe the effect of two-body annihilations and elastic scattering processes
with non-DM particles $\psi$, respectively.
Here $H=\dot a/a$ is the Hubble parameter, with $a$ the Friedman-Robertson-Walker scale factor,
${\left|\mathcal{M}\right|}^2$ is the squared amplitude summed over {\it both} initial and final state
internal or spin degrees of freedom, $g_i$, 
and for ease of notation we suppressed the implicit sum over all contributing 
particle species $\psi$. We also introduced $\bar f_i\equiv1-\varepsilon_i f_i$ to capture the effect 
of final-state Pauli blocking for fermions ($\varepsilon_{\chi,\psi}=+1$) and Bose enhancement for Bosons 
($\varepsilon_{\chi,\psi}=-1$).

For standard freeze-out calculations, it is assumed that all relevant particle species $\psi$ are
in full thermal equilibrium with the standard model, such that 
$f_\psi(\omega)=1/\left[\exp(\omega/T)\pm1 \right]$  at a photon temperature $T$.
Assuming furthermore that the DM particles are non-relativistic and stay in kinetic equilibrium 
with $\psi$ during the entire freeze-out process, Eq.~(\ref{diff_boltzmann}) can then be
integrated~\cite{Gondolo:1990dk} to an evolution equation for the DM number density $n_\chi$ that takes 
the familiar form
\be
\label{eq:GG}
\dot n_\chi +3H n_\chi=-\left\langle\sigma v\right\rangle\left(n_\chi^2-n_{\chi, {\rm MB}}^2\right)\,.
\ee
If the total entropy $s$ is conserved, this can equivalently be stated as an equation for the DM 
abundance $Y\equiv n_\chi/s$:
\be
\label{eq:dYdx}
\frac{dY}{dx}=\frac{\langle \sigma v\rangle}{x s \tilde H}\left(n_\chi^2-n_{\chi, {\rm MB}}^2\right)\,.
\ee
In the above equations, 
$n_\chi^{\rm MB}\equiv g_\chi(2\pi)^{-3}\int d^3p\,f_\chi^{\rm MB}=g_\chi m_\chi^2 T K_2(x)/(2\pi^2)$
denotes the number density of DM particles following a Maxwell-Boltzmann distribution of temperature $T$, 
 with $x\equiv m_\chi/T$ and $K_2$ being a modified Bessel function of the second kind, 
and
\be
 \label{eq:svav_GG}
\left\langle\sigma v\right\rangle 
 =  \int_1^\infty\!\!\! d\tilde s\,
 \frac{4 x\sqrt{\tilde s}({\tilde s-1})\, K_1\!\left({2{\sqrt{\tilde s}} x}\right)}
 {{K_2}^2(x)}
  \sigma_{\chi\chi\rightarrow \psi\psi}
\equiv \int_1^\infty\!\!\! d\tilde s\,
 \frac{x\sqrt{\tilde s-1}\, K_1\!\left({2{\sqrt{\tilde s}} x}\right)}
 {2m_\chi^2  {K_2}^2(x)}
  W_{\rm eff}(s)
\ee
is the thermally averaged annihilation cross section w.r.t.~such a distribution. Further,
$\tilde H\equiv H/\left[1+ (1/3)d(\log g_*)/d(\log T)\right]$, where the 
number of entropy degrees of freedom $g_*$ is defined through the relation 
$s\equiv (2\pi^2/45)g_* T^3$. $W_{\rm eff}$ is referred to as the (effective) invariant rate.

Eqs.~(\ref{eq:GG}--\ref{eq:svav_GG}) are widely used in relic density calculations,
not the least because they take the same form even when including co-annihilations~\cite{Edsjo:1997bg}.
These equations have been implemented to a high numerical precision since the 
early versions of \ds, and we refer to Refs.~\cite{ds4,Bringmann:2018lay} for a detailed
description. Notable additions since \ds~6.1 include updated degrees-of-freedom tables,
based on lattice simulations as well as perturbative  computations up to the 3-loop 
level~\cite{Laine:2006cp,Laine:2015kra}, and a major revision of 
how $\sigma_{\chi\chi\rightarrow \psi\psi}$ in Eq.~(\ref{eq:svav_GG}) is tabulated while solving Eq.~(\ref{eq:dYdx}).  
In detail this method uses an adaptive integrator to on the fly 
tabulate $W_{\rm eff}$ where needed; if nearby points have already been tabulated, an interpolation 
is used instead of the numerically expensive calculation of $W_{\rm eff}$. On top of this, before this tabulation starts, possible 
resonances in $W_{\rm eff}$ are inspected to see if they can be accurately fit with a Breit-Wigner form. If this is the case, the 
analytic Breit-Wigner form will be used instead of the actual expensive calculation of $W_{\rm eff}$. Compared to earlier 
tabulation methods in DarkSUSY, this new method is typically both faster and more accurate. 

A further addition in \ds\ 6.3 is that the Hubble expansion rate now is a replaceable function, 
\code{dsrdHubble}, allowing e.g.~for relic density calculations in cosmologies with non-standard expansion histories.

In the following we describe in some more detail recent code updates that allow relic density
calculations also in situations where one or more of the assumptions leading to 
Eqs.~(\ref{eq:GG}--\ref{eq:svav_GG})
are not satisfied. This has highly relevant applications for example when the DM particles are part of a 
secluded dark sector (section \ref{sec:dsfo}), the relic density is dominantly set by annihilation through 
a narrow resonance  (section \ref{sec:cbe}) or for freeze-in production of feably interacting DM
particles (section \ref{sec:fi}).

\subsection{Dark sector freeze-out}
\label{sec:dsfo}

Secluded dark 
sectors~\cite{Pospelov:2007mp,Pospelov:2008zw,Feng:2008mu}
constitute a prominent class of models where the DM relic density is not as usual set by freeze-out
from the SM heat bath.  The underlying idea of such scenarios is that DM might be interacting very weakly with 
the SM, but still sufficiently strongly to equilibrize with itself or other new particles. In general one therefore has to 
distinguish the photon
temperature, $T$, from that of the DM (and other new `dark') particles, $T_\chi$. For example, the visible and 
the dark sector could have been in thermal contact at high temperatures, with $T_\chi=T$, and only later
decoupled at some temperature $T_{\rm dec}$. In that case, since entropy is typically conserved separately 
in the two sectors, the temperature ratio will evolve as
\be
\label{xi_eq}
 \xi(T)\equiv \frac{T_\chi(T)}{T}=
 \frac{\left[{g_*^\mathrm{SM}}(T)/ {g_*^\mathrm{SM}}(T_\mathrm{dec}) \right]^\frac13}
 {\left[{g_*^\mathrm{DS}(T)}/{g_*^\mathrm{DS}(T_\mathrm{dec})}\right]^\frac13}\,,
 \ee
where $g_*^\mathrm{SM,DS}$ refers to the effective number of relativistic entropy d.o.f.~in the visible 
and dark sector, respectively.\footnote{%
It is worth noting that this relation tacitly assumes that at least one of the additional particles $\psi$ that the DM 
particles interact with has vanishing chemical potential. If all other dark sector particles are massive, too, this is in 
general no longer the case. See Ref.~\cite{Bringmann:2020mgx} for how to treat such situations.
}

In order to accurately describe the freeze-out of DM particles from such a secluded dark sector, 
the standard Boltzmann equation \eqref{eq:GG} must be adapted at three places: 
\begin{enumerate}
\item the equilibrium density $n_{\chi, {\rm MB}}$ must be evaluated at $T_\chi$ rather than the SM 
temperature $T$;
\item the same replacement, $x\to x/\xi$, must also be made in the expression for $\langle\sigma v\rangle$ 
in Eq.~(\ref{eq:svav_GG});
\item the additional energy content of the dark sector must be reflected in an increased Hubble
rate. During radiation domination, in particular, this implies $H^2=(8\pi^3/90)g_{\rm tot}M_{\rm Pl}^{-2}T^4$, 
with $g_{\rm tot}= g_{\rm SM} + \xi^4 g_{\rm DS}$ and $g_{\rm SM, DS}$ denoting the number of 
relativistic {\it energy} d.o.f.~in the visible and dark sector, respectively.
\end{enumerate}

The option to perform such relic density calculations has been implemented in \ds\ in the context of
a more general endeavour~\cite{Bringmann:2020mgx} to update precision calculations of the `thermal'
annihilation cross section, i.e.~the size of $\langle \sigma v\rangle$ that is needed -- close to freeze-out --  in
order to match the observed DM relic abundance, and how the presence of dark sectors affects 
the numerical value of this quantity. Concretely, two new functions \code{dsrddofDS} and \code{dsrdxi}
have been introduced and are now consistently used in all freeze-out calculations, returning 
$g_{\rm DS}(T_\chi)$ and $\xi(T)$, respectively.
If any of these functions is declared in a particle module, or when linking a main program, this automatically
replaces their trivial implementation in the \code{main} library that corresponds to standard freeze-out 
(i.e.~$g_{\rm DS}\equiv0,\xi\equiv1$). In order to facilitate the implementation of such a model-specific 
version of \code{dsrddofDS}, \ds\ furthermore provides an auxiliary function \code{dsrdsingledof} that
returns the temperature-dependent effective number of relativistic degrees of freedom for a single 
massive particle (resulting in exactly $1$ and $7/8$ for bosons and fermions, respectively, in the massless limit).
An example of \code{dsrdxi} implementing Eq.~(\ref{xi_eq}) can be found in the \code{vdSIDM} module.

\subsection{Freeze-out beyond kinetic equilibrium}
\label{sec:cbe}

The standard treatment of the freeze-out process consists in numerically solving Eq.~(\ref{eq:GG})
-- which rests, just like the extensions described in section \ref{sec:dsfo}, on the assumption
that DM remains in {\it kinetic} equilibrium during freeze-out.\footnote{%
\label{foot:kineq}
Technically, what enters in the derivation~\cite{Gondolo:1990dk} of Eqs.~(\ref{eq:GG}--\ref{eq:svav_GG}) is that
the DM phase-space distribution is of the form $f_\chi(p,T)=A(T)\exp(-E/T)$
both before and during the entire decoupling process.
}
There is however an important subset of DM models that feature a strongly velocity-dependent annihilation
cross section and where this assumption is not met. For example, using the conventional Eq.~(\ref{eq:GG}) 
may result in a relic density that does not even have the correct order
of magnitude in models with narrow resonances 
(like in particular for the standard model Higgs boson), Sommerfeld-enhanced annihilation, or for DM particles degenerate
in mass with the annihilation products~\cite{vandenAarssen:2012ag,Duch:2017nbe,Binder:2017rgn,Binder:2017lkj,Binder:2021bmg}.
In such situations, a convenient alternative to the numerically challenging integration of the full
Boltzmann equation at the phase-space level, Eq.~(\ref{diff_boltzmann}),  is to consider a set of Boltzmann
equations that couple the evolution of the DM number density and velocity dispersion, respectively
(as first introduced in Ref.~\cite{vandenAarssen:2012ag}, and later refined in Ref.~\cite{Binder:2017rgn}).

Introducing $T_\chi\equiv g_\chi/(3n_\chi)\int d^3p\,(2\pi)^{-3} (p^2/E) f_\chi$ and $y(x)\equiv m_\chi T_\chi s^{-2/3}$,
in analogy to $n_\chi$ and $Y=n_\chi/s$, these equations generalize Eq.~(\ref{eq:dYdx}) to
\bea
\frac{x}{Y}\frac{dY}{dx} &=& \frac{s Y}{\tilde H}\left[
\frac{Y_{\rm eq}^2}{Y^2} \left\langle \sigma v\right\rangle_T- \left\langle \sigma v\right\rangle_{T_\chi}
\right]\,, \label{Yfinalfinal}\\
\frac{x}{y}\frac{dy}{dx} &=& \frac{\gamma w}{\tilde H}\left[\frac{y_{{\rm eq}}}{y} -1\right]
+\frac{sY}{\tilde H}\left[
\left\langle \sigma v\right\rangle_{T_\chi}-\left\langle \sigma v\right\rangle_{2,T_\chi}
\right] \label{yfinalfinal} 
+\frac{sY}{\tilde H}\frac{Y_{\rm eq}^2}{Y^2}\left[
\frac{y_{{\rm eq}}}{y}\left\langle \sigma v\right\rangle_{2,T}\!-\!\left\langle \sigma v\right\rangle_T
\right]
+2(1-w)\frac{H}{\tilde H}\,.\nonumber\\
\eea
Here, a subscript $T$ or $T_\chi$ indicates the temperature at which to take thermal averages, and
$\langle \sigma v\rangle_{2}$ is a variant of Eq.~(\ref{eq:svav_GG}) that is explicitly stated in 
Ref.~\cite{Binder:2017rgn}; the parameter $w(T_\chi)\equiv 1-{\langle p^4/E^3 \rangle_{T_\chi}}/({6T_\chi})$ 
indicates deviations from DM being highly non-relativistic (where $w=1$). The momentum transfer rate $\gamma(T)$,
finally, is given by
\begin{align}
  \label{cTdef}
 \gamma= \! \frac{1}{3 g_{\chi} m_{\chi} T} \! \int \! \frac{\text{d}^3 k}{(2\pi)^3} g^{\pm}(\omega)\left[1\!\mp\! g^{\pm}(\omega)\right] \! \! \! \int\limits^0_{-4 k_\mathrm{cm}^2} \! \! \!  \text{d}t (-t) \frac{\text{d}\sigma}{\text{d}t} v\,,
\end{align}
where ${k}_\mathrm{cm}^2  \equiv m_\chi^2{k}^2/(m_\chi^2+2\omega m_\chi+m_f^2)$ and $ |\mathcal{M}|^2$ 
in $({\text{d}\sigma}/{\text{d}t}) v \equiv |\mathcal{M}|^2_{\chi f\leftrightarrow\chi f}/(64 \pi {k} \omega m_\chi^2)$
is evaluated at $s\simeq m_\chi^2+2\omega m_\chi+m_f^2$. By construction, these equations accurately describe
the evolution of $Y(x)$ as long as efficient DM self-interactions force $f_\chi(t,p)$ into a thermal shape with
$T_\chi\neq T$ (or, rather, as long as the resulting thermal averages are not significantly affected by deviations from
such a thermal shape); even more generally, in fact, relic density calculations based on these equations
often provide a good estimate of the relic density that results from directly integrating Eq.~(\ref{diff_boltzmann}).
For a more detailed discussion see Ref.~\cite{Binder:2021bmg}.

The numerical solution of the coupled system of Eqs.~(\ref{Yfinalfinal},\ref{yfinalfinal}) has been implemented in \ds\ 
in the context of Ref.~\cite{Binder:2021bmg}. Concretely, a routine \code{dsrdomega\_cBE} has been added that
returns the final DM relic density based on these coupled Boltzmann equations -- just as the conventional \code{dsrdomega}
returns the relic density based on a solution of Eq.~(\ref{eq:GG}). The usage of both routines is illustrated in detail
in a number of example programs located at \code{examples/aux/oh2\_*.f}. A call to \code{dsrdomega\_cBE} also
initializes the function \code{dsrdthav\_select} which for convenience returns the various thermal averages appearing in 
Eqs.~(\ref{Yfinalfinal},\ref{yfinalfinal}), including the momentum exchange rate $\gamma$. In order to enable a main program 
to use \code{dsrdomega\_cBE}, finally, a particle module must provide the typical interface functions required by 
{\it both} relic density and kinetic decoupling routines, i.e.~\code{dsrdparticles}, \code{dsanwx}, \code{dskdparticles} and
\code{dskdm2} (see the manual or Ref.~\cite{Bringmann:2018lay} for more details).

\subsection{Freeze-in}
\label{sec:fi}

Another important exception to the validity of  Eqs.~(\ref{eq:GG}) are DM particles with interactions so weak that
they never thermalized with the heat bath. Such feably interacting massive particles (FIMPs) could still
obtain the correct relic density through continuous production from the thermal bath of standard model particles,
a mechanism known as {\it freeze-in production} of DM~\cite{Hall:2009bx,Chu:2011be}. Starting from Eq.~(\ref{diff_boltzmann}),
the main technical difference to the freeze-out case is that a much larger range of temperatures are relevant 
for determining the final relic abundance, implying in particular that the DM particles can no longer be assumed to be
non-relativistic and that the effect of quantum statistics ($\bar f_i\neq1$), but also other thermal effects, potentially become relevant.
Still, it is possible to capture all these effects with a description that closely resembles that of the freeze-out 
case~\cite{Bringmann:2021sth}.

As long as the FIMP abundance stays far below the equilibrium abundance, in particular, it increases as
\be
\frac{dY_\chi}{dx}=\frac{n_{\chi, {\rm MB}}^2}{x s \tilde H}\langle \sigma v\rangle\,.
\label{eq:dYdx_fimp}
\ee
While this is {\it formally} the same as Eq.~(\ref{eq:dYdx}), without the `backreaction' term that would describe the annihilation 
of FIMPs into standard model particles, there are some important differences:
\begin{enumerate}
\item The term $n_\chi^{\rm MB}$ in Eq.~(\ref{eq:dYdx_fimp}) refers to the number density of a {\it would-be} 
Maxwell-Boltzmann distribution of DM particles. Unlike for WIMPs, cf.~footnote \ref{foot:kineq},
the above formulation does not assume in any way that the {\it actual}
DM distribution is related to a thermal one.
\item Just as in the case of WIMPs, Eq.~(\ref{eq:dYdx_fimp}) is formulated in terms of the DM
{\it annihilation} cross section $\sigma$.  The thermal effects appearing due to the presence of relativistic DM particles,
however, require a generalization of the thermal average given in Eq.~(\ref{eq:svav_GG}).
\end{enumerate}
Conveniently, the quantity $\langle\sigma v\rangle$ can still be expressed in terms of 
the same model-independent thermal kernel as in Eq.~(\ref{eq:svav_GG}). The model-dependent invariant 
rate $W_{\rm eff}$, on the other hand, now is also temperature-dependent and in general  given by
\be
\label{eq:weffT}
  W_{\rm eff}(s,T)\equiv
  16m_\chi^2 \frac{x\tilde s \sqrt{\tilde s-1}}{K_1\!\left({2{\sqrt{\tilde s}} x}\right)}
   \int_1^\infty d\gamma\, \sqrt{\gamma^2-1}e^{-2\sqrt{\tilde{s}} x \gamma}
    \sum_{\psi_1\psi_2}
   \sigma_{\chi\chi\to\psi_1\psi_2} (s,\gamma)\,.
\ee
Here the integration is over Lorentz boosts $\gamma$ from the center-of-mass to the cosmic
rest frame, and the in-medium cross section can be written as
\be
\label{eq:sigfinal_full}
\sigma_{\chi\chi\to\psi_1\psi_2} (s,\gamma)=
\frac{N_\psi^{-1}}{8\pi s}\frac{|\mathbf{k}_{\rm CM}|}{\sqrt{s-4m_\chi^2}}
\int_{-1}^{1}\frac{d\cos\theta}{2}\left|\overline{\mathcal{M}}\right|^2_{\chi\chi\to\psi_1\psi_2} \!(s,\cos\theta)\, G_{\psi_1\psi_2}(\gamma, s, \cos\theta)\,,
\ee
with $N_\psi=2$ for identical SM particles ($\psi_1=\psi_2$) and  $N_\psi=1$ 
otherwise. The quantity $G_{\psi_1\psi_2}(\gamma, s, \cos\theta)$ is explicitly stated in Ref.~\cite{Bringmann:2021sth}
and encodes the effect of quantum statistics in the final state, leading to an enhancement ($G_{\psi_1\psi_2}>1$) or 
decrease ($G_{\psi_1\psi_2}<1$) of
the corresponding cross section in vacuum; for $G_{\psi_1\psi_2}=1$, in particular, the definition
of $W_{\rm eff}(s,T)$ in Eq.~(\ref{eq:weffT}) becomes identical to that of the conventional invariant rate 
given in Eq.~(\ref{eq:svav_GG}). It is worth noting that a $T$-dependent $W_{\rm eff}$ also allows to include 
temperature-dependent effects other than those due to quantum statistics, such as thermal masses and phase
transitions, which can affect both interaction rates and the spectrum of relevant SM states.

The capability of \ds\ to perform freeze-in calculations has been added in the context of Ref.~\cite{Bringmann:2021sth},
which discusses in detail freeze-in production of Scalar Singlet DM~\cite{Silveira:1985rk}.
In particular, \code{dsfi2to2oh2} numerically solves Eq.~(\ref{eq:dYdx_fimp}) and 
returns the resulting DM relic density, $\Omega_\chi h^2$, as a function of the reheating
temperature (defined as the starting point of the integration). During the numerical integration, special care 
is taken to model the effects of QCD and EW phase transitions to sufficient accuracy. The thermally averaged cross 
section is provided by the function \code{dsfithav} (rather than \code{dsrdthav} as in 
the WIMP case). There are two interface functions that a particle physics module must provide for the 
freeze-in routines in the \ds\ \code{core} library to work: a subroutine \code{dsrdparticles}, providing kinematic information
about (potentially $T$-dependent) thresholds and resonances, and a function  \code{dsanwx\_finiteT} returning
the temperature-dependent effective invariant rate as defined in Eq.~(\ref{eq:weffT}).
Two example programs,
\code{examples/aux/FreezeIn\_ScalarSinglet} and \code{examples/aux/FreezeIn\_generic\_fimp}, illustrate
the usage of the freeze-in routines for the Scalar Singlet model and a `generic' FIMP model, 
respectively (where the latter implements a simplified contact-like interaction with 
$\left|\mathcal{M}\right|^2\equiv c\left({s}/{\Lambda^2} \right)^n$; for 
further details we refer to the manual and Ref.~\cite{Bringmann:2021sth}). As of version 6.3, \ds\ also provides
temperature-dependent SM masses (\code{dsmass\_finiteT}) and Higgs vev (\code{dshvev\_finiteT}), as well as
an improved treatment of the partial Higgs decay width including in particular hadronic final states (\code{dssmgammahpartial.f}) 
-- all of which must be modelled accurately e.g.~for freeze-in calculations involving Higgs portal models~\cite{Bringmann:2021sth}.

\section{Direct detection: cosmic-ray upscattering of dark matter}
\label{sec:crdm}

Conventional direct detection experiments were long thought to be insensitive to sub-GeV DM particles~\cite{XENON:2018voc},
because the typical kinetic energy of Galactic DM is too small to trigger nuclear recoil energies 
above the necessary threshold. For large elastic scattering cross sections with nuclei, however,
there inevitably exists an irreducible flux of relativistic DM particles that are up-scattered by high-energy
cosmic rays~\cite{Bringmann:2018cvk, Cappiello:2018hsu}. This sub-dominant component of the expected DM
flux at Earth allows to constrain both very light DM particles and DM particles in the GeV range 
that would otherwise be stopped in the overburden before reaching the 
detector~\cite{Ema:2018bih,Cappiello:2019qsw,Dent:2019krz,Xia:2020apm,Xia:2021vbz,CRDMupdate}.

The local interstellar flux of such cosmic-ray upscattered DM (CRDM) particles is given by
\be
\frac{d\Phi_{\chi}}{dT_\chi}=
D_\mathrm{eff} \frac{\rho_\chi^\mathrm{local}}{m_\chi}  
\sum_N
\int_{T_N^\mathrm{min}}^\infty d T_N\, \frac{d \sigma_{\chi N} }{dT_\chi} \frac{{d\Phi^\mathrm{LIS}_N}}{dT_N}
\label{eq:chiCR}
\,,
\ee
where ${d \sigma_{\chi N} }/{dT_\chi}$ is the differential elastic scattering cross section
to accelerate DM to a kinetic recoil energy of $T_\chi$, for an incident cosmic-ray (CR) nucleus $N$ with energy $T_N$,
and ${{d\Phi_N}^{\rm LIS}}/{dT_N}$ is the {\it local} interstellar CR flux; $\rho_\chi$ is the local DM density
and $D_{\rm eff}\sim10$\,kpc is an effective distance out to which the above expression for the production of
the CRDM component holds (for further details, see Refs.~\cite{Xia:2021vbz,CRDMupdate}).
The scattering rate of relativistic DM particles in underground detectors is formally given by the same expression
as for the standard non-relativistic contribution, i.e.
\be
\label{eq:gammarate}
 {\frac{d\Gamma_N}{d T_{N}}=
 \int_{T_\chi^{\rm min}}^\infty \!\!dT_\chi\ 
 \frac{d \sigma_{\chi N}}{dT_N} \frac{d\Phi_\chi}{dT_\chi}} \,,
\ee
with $T_\chi^{\rm min}$ being the minimal DM energy needed in order to induce a nuclear recoil $T_N$.
Here, the nuclear scattering cross section is in general a function of both the center-of-mass energy and 
the (spatial) momentum transfer, $Q^2=2m_N T_N$.
If the dominant dependence on $Q^2$ factorizes -- like in particular 
for form factors -- the above rate has an identical $Q^2$-dependence for relativistic and non-relativistic DM.
It is then straight-forward to re-interpret published conventional direct detection results into limits on the CRDM 
component (the same also applies to neutrino detectors sensitive to nuclear recoils, after converting the nuclear recoil to 
the detected apparent electron energy $T_e$)~\cite{Bringmann:2018cvk}.

For large scattering cross sections the complication arises that the CRDM flux in Eq.~(\ref{eq:chiCR}) is not 
necessarily the one that is relevant for underground laboratories, because the original CRDM flux 
is attenuated by scattering with nuclei in the overburden of the experimental location. In other words, the kinetic energy $T_\chi^z$
of a DM particle at depth $z$ of the detector may  be significantly less than its 
initial energy $T_\chi$ at the top of the atmosphere ($z=0$). The expression for the rate in Eq.~(\ref{eq:gammarate}) 
then continues to apply, after a change of variables from $T_\chi^z$ to $T_\chi(T_\chi^z)$, but the scattering cross section 
${d \sigma_{\chi N}}/{dT_N} $ must still be evaluated at the actual DM energy $T_\chi^z$ at the detector location.
In order to find the {\it average} kinetic energy at the detector location, one needs to solve the energy loss equation
\be
\label{eq:eloss}
\frac{dT_\chi^z}{dz}=-\sum_N n_N\int_0^{T_N^\mathrm{max}}\!\!\!dT_N\,\frac{d \sigma_{\chi N}}{dT_N} T_N\,,
\ee
where the sum runs over the most dominant nuclei in the overburden, i.e.~no longer over the CR species as in Eq.~(\ref{eq:chiCR}).
When expressed in terms of an integration over $Q^2$ instead of $T_N$, the above expression can 
also be used to include the attenuation due to {\it inelastic} scattering, which dominates at high momentum transfers; 
to a good approximation, this can be modelled
by adding a model-independent function $I^2(Q^2)$ to the usual nuclear form factors~\cite{CRDMupdate}.

The necessary routines to compute DM limits resulting from the irreducible CRDM flux have been implemented and 
released with version 6.2 of \ds\ in the context 
of Ref.~\cite{Bringmann:2018cvk}, with significant additions (full $Q^2$ and $s$-dependent elastic scattering cross sections,
inelastic scattering, increased number of nuclei contributing to scattering processes) added 
subsequently~\cite{Bondarenko:2019vrb,CRDMupdate}. The interstellar CRDM flux, Eq.~(\ref{eq:chiCR}), is returned
by the function \code{dsddDMCRflux}, based on the dominant species in the CR flux 
${{d\Phi^\mathrm{LIS}_N}}/{dT_N}$~\cite{Boschini:2018baj,Boschini:2020jty} (provided by \code{dscrISRflux}).
The full relativistic scattering cross sections $d\sigma_{\chi N}/dT_N$ and $d\sigma_{\chi N}/dT_\chi$ that appear
in the above expressions are implemented as the conventional cross sections in the highly non-relativistic limit 
-- including state-of-the-art nuclear form factors -- and then multiplied by a relativistic correction factor 
\code{dsddsigmarel} to take into account the model-dependent dependence on $s$ and $Q^2$; in order to 
use the CDMR routines for arbitrary scattering cross sections, one thus only has to replace the function
\code{dsddsigmarel}.
While there is a separate function to calculate the rate in Eq.~(\ref{eq:gammarate}), namely \code{dsddDMCRdgammadt}
directly, it is in practice most convenient to call the `driver' routine \code{dsddDMCRcountrate} which only takes
the name of a given -- direct detection or neutrino -- experiment as input and directly returns the experimental count rate 
from the CRDM component,
divided by the rate corresponding to the published limit of that experiment; this ensures that all experiment-specific 
settings are initialized correctly
(for example the depth of the detector location, and the composition of the material in the overburden).
The usage of  \code{dsddDMCRcountrate} is demonstrated in the example program \code{examples/aux/DDCR\_limits.f} 
(while \code{examples/aux/DDCR\_flux.f} provides an illustration of how to compute and tabulate the interstellar CRDM flux).

\section{Indirect detection: particle yields}
\label{sec:id}

The particle yields from DM annihilation or decay in the halo and in the Sun/Earth have traditionally, in \ds, been generated 
with {\sf Pythia} 6~\cite{Sjostrand:2006za}. 
However, to allow for more flexibility and improved yield calculations, the possibility to use tables 
from {\sf Pythia} 8~\cite{Sjostrand:2014zea} runs is being added in \ds\ 6.3, based on  {\sf Pythia} 8.306 with default settings. 
For annihilations in the Sun/Earth, WimpSim~\cite{Edsjo:2007ns} is used to handle hadron interactions and neutrino oscillations, 
and has correspondingly been updated to be based on {\sf Pythia} 8 as an event generator as well. 

For the {\sf Pythia} 8 simulations we include for a range of 
(annihilating) DM masses between 5 GeV and 20 TeV. We use a total of 30 different masses and simulate annihilations into 15 different 
final states (all quark-antiquark final states, glue-glue, $W^+W^-$, $Z^0Z^0$, $\tau^+ \tau^-$, prompt neutrino final states, $hh$ and 
$Z^0 h$, with $h$ being the standard model Higgs boson). For each mass and annihilation channel, $10^7$ annihilation events have 
been simulated. For neutrino oscillations, we use the NuFit 5.1 normal ordering best fit values \cite{Esteban:2020cvm,nufitonline}. For 
interactions at a detector on Earth, simulations were performed for IceCube during the austral winter, but the results are to within a few 
percent applicable also for other detectors. The final yields are tabulated in both energy and angle, and then interpolated via the 
routines in \ds. 
Compared to the {\sf Pythia} 6 runs, the {\sf Pythia} 8 runs currently have lower statistics and hence the {\sf Pythia} 6 runs will 
initially still be kept as the default. Eventually, with higher statistics runs becoming available, the default will 
change to yield tables based on {\sf Pythia} 8. Independently of the default settings, the user can easily change which yield tables 
to use by calling \code{dsseyield\_set}.

Similarly, for indirect DM detection rates related to annihilations or decays in the halo, we also rely on event generators to calculate the 
yield of positrons, antiprotons, gamma rays, neutrinos and anti-deuterons for a range of different DM masses and annihilation/decay 
channels. The traditional high-statistics tables based on {\sf Pythia} 6 runs remain the default here, but we are in the 
process of adding {\sf Pythia} 8 tables also for these yields, where we use the same range of masses and set of annihilation channels as
described above for annihilation in the Sun/Earth (except that we here also include annihilation into $\mu^+ \mu^-$). 

On top of this, \ds\ now allows to 
use alternative yield calculations in its indirect detection routines; currently, this includes yield tables provided 
by Refs.~\cite{Amoroso:2018qga,Plehn:2019jeo,Bauer:2020jay,Jueid:2022qjg}.
The main difference between the various implemented yield tables concerns statistics, masses and channels that are simulated, but also 
assumptions regarding the underlying physics. Concretely, the (current) default tables based on {\sf Pythia} 6 runs with very high statistics 
are particularly well tested and also include anti-deuteron yields; corresponding tables based on {\sf Pythia} 8 runs, with default settings,
will soon be available -- but initially ship with somewhat reduced statistics compared to the present \ds\  implementation, 
and include all final 
states except anti-deuterons in the initial release. Externally provided yield tables are best described in the respective
references. Roughly speaking, the tables by Bauer {\it et al.}~\cite{Bauer:2020jay} focus specifically on accurately modelling the 
yield in the multi-TeV regime and beyond,
where {\sf Pythia} ceases to be reliable, while those by Plehn {\it et al.}~\cite{Plehn:2019jeo} focus on the sub-GeV regime
and the impact of hadronic resonances that are not (fully) included in {\sf Pythia}, either. The yield tables by 
Jueid {\it et al.}~\cite{Jueid:2022qjg} (as well as an earlier version by Amoroso et 
al.~\cite{Amoroso:2018qga}), on the other hand, are also based on {\sf Pythia} 8 and cover a similar DM mass range as in the  
\ds\ default implementation -- but are based on a number of different treatments of the underlying QCD uncertainties related to the
fragmentation of light quarks; the present \ds\ implementation returns their central prediction for the spectra.

Accessing the yield routines is straight-forward, as demonstrated in the example program  
\code{examples/aux/wimpyields.f}. In particular, a simple initial call to \code{dsanyield\_set} allows
to switch between the implemented yield tables, as well as to choose between various options
(e.g.~SM-like or $B-L$-like models for the tables from Plehn {\it et al.}~\cite{Plehn:2019jeo}). Such 
a call to \code{dsanyield\_set} does not only affect the output of \code{dsanyield\_sim}, as demonstrated
in this program, but automatically adjusts the output of all routines returning indirect detection rates 
(e.g.~gamma-ray or positron fluxes) as well.

\section{Summary}
\label{sec:sum}

\ds\ is a versatile numerical tool to compute potential DM observables that is widely used in the community.
Here we have presented the most important updates since version 6.1~of the code, ranging
from relic density computations beyond the standard thermal equilibrium assumption (section \ref{sec:rd}) 
to direct detection routines including the effect of cosmic-ray upscattering
(section \ref{sec:crdm}) and various versions of updated yield routines relevant for indirect detection
(section \ref{sec:id}). \ds\ continues to be in a state of active development, and 
the range of possible applications is expected to further increase in the near future.
We thus recommend to regularly check {\tt  www.darksusy.org} for new releases,
updated documentation, as well as new concrete example applications to get started. Using the code is simple
and straightforward (see also Appendix \ref{app}), but we happily encourage users to get in touch with 
the developers in case of problems -- as well as for suggestions concerning further code additions or
improvements.

\acknowledgments

We warmly thank Lars Bergstr\"om, Paolo Gondolo and Piero Ullio for their invaluable help in bringing \ds\ 6 into its present 
shape. The development of \ds\ since version 6.1 has also greatly benefitted from direct code 
contributions by collaborators on various physics projects; in this context we are particularly grateful to Saniya Heeba, 
Felix Kahlhoefer, Helena Kolesova and Kristian Vangsnes. We further wish to thank all users who
provided feedback that helped to improve the code and make it more system-independent.

\appendix
\section{Technical updates: installation and make system}
\label{app}

The installation of \ds\ is in principle as straightforward as downloading the most recent version
from {\tt www.darksusy.org/download.html}, unpacking the \code{darksusy-6.x.x.tgz} file and running
\begin{verbatim}
./configure
make
\end{verbatim}
in the newly created directory \code{darksusy-6.x.x/}. While typically not necessary, special options can as usual
be specified in the \code{configure} step, for example to choose a specific compiler and compiler-specific 
performance flags (see, e.g., the script \code{./conf.gfortran} for an example optimized for the use with \code{gfortran}). 
Assuming that several cores are available, furthermore, the make command can typically be made significantly faster by adding 
the \code{-j} flag (which, however, makes the output harder to follow). 
 
As indicated by the terminal output, the \code{make} command first builds {\it contributed} code, i.e.~external 
numerical packages that a subset of \ds's routines relies on,\footnote{%
Presently, these include {\sf HEALPix}~\cite{Gorski:2004by},  {\sf FeynHiggs}~\cite{Heinemeyer:1998yj,Hahn:2009zz}, {\sf HiggsBounds}~\cite{Bechtle:2013wla}, {\sf HiggsSignals}~\cite{Bechtle:2013xfa},
{\sf ISAJET}~\cite{Paige:2003mg} and {\sf SuperIso}~\cite{Mahmoudi:2008tp}.
}
 before proceeding to compile genuine \ds\ code (resulting in the main library, 
\code{lib/libds\_core.a}, as well as a separate library for each of the shipped particle modules). 
The compilation of those external libraries is included for convenience, not the least to ensure a seamless interface with the
rest of the code, but sometimes more demanding in terms of system requirements than that of the proper \ds\ libraries. 
The problem with this setup used to be that failing to compile contributed code would also stop the installation process for the
rest of \ds. The updated \code{make} system takes care of this by just displaying a warning in such a situation and then,
if possible with minimal damage, automatically removing any dependence of \ds\ on the respective contributed code
package. If {\sf HEALPix} fails to compile, e.g., the \ds\ main and module libraries will still be built without any problems, 
the only compromise being that a call to the integration routine \code{dshealpixint} will result in a corresponding warning
message (for an example of {\sf HEALPIx}-based line-of-sight integrations of the astrophysical $J$- or $D$-factors,
another newly added feature, see the short demonstration program \code{examples/aux/DMhalo\_los}).
Likewise, the only noticeable impact of failing to compile {\sf HiggsBounds} or {\sf HiggsSignals} is that a call
to \code{dshiggsbounds} will not as usual compute the $p$-value based on Higgs observables, in the \code{mssm} module,
but instead return a warning that support for these libraries is disabled. 

For compilation problems beyond the cases that can be handled automatically,
furthermore, a new simplified target is available. The sequence of calls
\begin{verbatim}
make distclean
./configure
make darksusy_light
\end{verbatim}
restores the pristine version of the code (i.e.~the one after downloading and unpacking), and then installs a {\it `light'}
version of \ds\ that does not depend on any contributed code at all. While this will disable particle modules that 
heavily rely on contributed code, in particular the \code{mssm} module, most of the functionality of the \code{core}
library as well as the majority of the particle modules will not be affected and build as usual.

Finally it is worth mentioning that the result of the standard installation process, as described above, consists of 
{\it static} libraries located in \code{lib/}. 
Interfacing \ds\ with other codes, however, sometimes requires {\it shared} libraries instead. We therefore now also 
provide corresponding targets, e.g.~\code{make ds\_mssm\_shared}, which are heavily used for 
example in GAMBIT~\cite{GAMBIT:2017yxo,Bloor:2021gtp}.

\bibliographystyle{JHEP_improved}
\bibliography{biblio.bib}

\end{document}